\documentclass[10pt,conference]{IEEEtran}\IEEEoverridecommandlockouts
\usepackage{epsfig,graphicx,subfigure,psfrag,amsmath,cases,bm}
\usepackage{latexsym,amssymb,amsmath,epsfig,subfigure,algorithm,mathtools}
\usepackage{algorithmic}
\usepackage{color}
\usepackage{url}
\usepackage{scrtime}
\usepackage[left=0.6in,top=0.40in,bottom=0.5in,right=0.6in]{geometry}
\author{\IEEEauthorblockN {Yan Sun, Derrick Wing Kwan Ng, Dongfang Xu, Linglong Dai, and Robert Schober\vspace*{-13mm}}

\thanks{Y. Sun, D. Xu, and R. Schober are with Friedrich-Alexander-University Erlangen-N\"urnberg (FAU), Germany. D. W. K. Ng is with University
of New South Wales, Australia. L. Dai is with Tsinghua University, Beijing, China.%
}
}

\title{Resource Allocation for Solar Powered UAV Communication Systems\vspace*{-1mm}}

\newtheorem{Thm}{Theorem}

\newtheorem{T-Prob}{Transformed Problem}

\DeclareMathOperator{\maxo}{maximize}
\DeclareMathOperator{\mino}{minimize}

 \newcommand{\qed}{\hfill \ensuremath{\blacksquare}}

\newcommand{\abs}[1]{\lvert#1\rvert}
\newcommand{\norm}[1]{\lVert#1\rVert}
\begin{document}
\IEEEspecialpapernotice{(Invited Paper)\vspace*{-0.5mm}}
\maketitle\vspace*{-2mm}

\begin{abstract}
In this paper, we investigate the resource allocation design for multicarrier (MC) systems employing a solar powered unmanned aerial vehicle (UAV) for providing communication services to multiple downlink users. We study the joint design of the three-dimensional positioning of the UAV and the power and subcarrier allocation for maximization of the system sum throughput. The algorithm design is formulated as a mixed-integer non-convex optimization problem, which requires a prohibitive computational complexity for obtaining the globally optimal solution. Therefore, a low-complexity suboptimal iterative solution based on successive convex approximation is proposed. Simulation results confirm that the proposed suboptimal algorithm achieves a substantially higher system sum throughput compared to several baseline schemes.
\end{abstract}
\renewcommand{\baselinestretch}{0.91}
\vspace*{-0mm}
\section{Introduction}
Future wireless communication systems are envisioned to provide ubiquitous and sustainable high data-rate communication services \cite{Andrews5GSurvey,book:Key5GWong}.
However, in some cases, deploying conventional terrestrial infrastructure (e.g. base stations (BSs)) is not cost-effective or not feasible. For example, it is not possible to deploy fixed BSs in a timely and economical manner in temporary hotspots, disaster areas, and complex terrains.
To handle this issue, aerial communication systems based on unmanned aerial vehicles (UAVs) have been proposed as a promising new paradigm to facilitate fast and flexible deployment due to their excellent maneuverability \cite{Zeng16Throughput}\nocite{wu2017joint,zhang2017securing,Yaliniz163DMaxUser}--\cite{Mozaffari16MaxCoverage}.
In particular, UAVs equipped with on-board wireless transmitters can fly over the target area and provide communication services.
Moreover, since UAVs enjoy high mobility, they can adjust their aerial position according to the real-time locations of the users which introduces additional degrees of freedom for improving system performance.
In \cite{Zeng16Throughput}, the authors investigated UAV trajectory design for minimization of the mission completion time in multicasting systems.
The authors of \cite{wu2017joint} proposed a suboptimal joint trajectory, power allocation, and user scheduling algorithm for maximization of the minimum user throughput in multi-UAV systems.
In \cite{zhang2017securing}, a suboptimal joint trajectory and power allocation algorithm was proposed for maximization of the system secrecy rate in a UAV communication system.
The three-dimensional (3-D) positioning of UAVs for maximization of the number of served users and the coverage area was studied in \cite{Yaliniz163DMaxUser} and \cite{Mozaffari16MaxCoverage}, respectively.
However, the UAV-based communication systems considered in \cite{Zeng16Throughput}\nocite{wu2017joint,zhang2017securing,Yaliniz163DMaxUser}--\cite{Mozaffari16MaxCoverage} are powered by on-board batteries, leading to limited operation time. Specifically, the UAVs in
\cite{Zeng16Throughput}\nocite{wu2017joint,zhang2017securing,Yaliniz163DMaxUser}--\cite{Mozaffari16MaxCoverage} are required to return to the ground frequently for recharging their batteries.
Hence, these designs cannot guarantee stable and sustainable communication services which creates a system performance bottleneck.

To overcome this shortcoming, solar powered UAVs have received significant attention due to their potential to realize perpetual flight \cite{oettershagen2016perpetual,morton2015solar}.
In particular, solar panels installed on the UAVs harvest the received solar energy and convert it to electrical energy for long endurance flights.
The authors of \cite{oettershagen2016perpetual} and \cite{morton2015solar} have developed solar powered UAV prototypes and demonstrated the possibility of continuous flight for $28$ hours.
However, the amount of harvested solar energy is affected by the flight altitude of the UAV.
For example, the atmospheric transmittance decreases for lower altitudes leading to a smaller amount of harvested solar energy \cite{duffie2013solar}.
Besides, the intensity of solar energy can significantly decrease if the light passes through clouds, resulting in reduced solar energy flux at the solar panels \cite{duffie2013solar,kokhanovsky2004optical}.
Therefore, UAVs flying at higher altitude can generally harvest more solar energy than those flying at lower altitude.
In \cite{Lee2017PathSolarUAV}, the authors studied the optimal trajectory of solar-powered UAVs for maximization of the harvest solar power taking into account the atmospheric transmittance.
However, the influence of clouds on solar powered UAVs was not considered in \cite{Lee2017PathSolarUAV}.
Moreover, \cite{Lee2017PathSolarUAV} focused only on flight control of solar powered UAVs, whereas communication design was not considered.
In fact, since higher flight altitudes lead to a more severe path loss for air-to-ground communications, there is a tradeoff between harvesting more solar energy and improving communication performance.
This tradeoff does not exist in conventional UAV communication systems  \cite{Zeng16Throughput}\nocite{wu2017joint,zhang2017securing,Yaliniz163DMaxUser}--\cite{Mozaffari16MaxCoverage} and the results derived in \cite{Zeng16Throughput}\nocite{wu2017joint,zhang2017securing,Yaliniz163DMaxUser}--\cite{Mozaffari16MaxCoverage} cannot be applied for solar powered UAV communication systems.
Moreover, multicarrier (MC) techniques are expected to play an important role also in future multiuser communication systems
\cite{Kwan13EEOFDMEnergyHarvest}\nocite{sun2016optimalJournal,Xiang17crosslayer,Zhu17Optimal,JR:WIPT_fullpaper,Bannour17MIMOOFDM,sun2017robustNOMA}--\cite{Zhou16Wireless}.
However, resource allocation designs for BS-based MC communication systems
\cite{Kwan13EEOFDMEnergyHarvest}\nocite{sun2016optimalJournal,Xiang17crosslayer,Zhu17Optimal,JR:WIPT_fullpaper,Bannour17MIMOOFDM,sun2017robustNOMA}--\cite{Zhou16Wireless}
cannot be directly applied to solar powered MC UAV communication systems, where the power and subcarrier allocation is coupled with the aerial positioning of the UAVs.
In fact, the joint design of the 3-D positioning and the power and subcarrier allocation for solar powered MC UAV communication systems is an open research problem.

In this paper, we address the above issues. To this end, the resource allocation algorithm design for solar powered MC UAV communication systems is formulated as a combinatorial non-convex optimization problem for maximization of the system sum throughput. The considered optimization problem is in general intractable and obtaining the globally optimal solution may result in prohibitive computational complexity. Therefore, we develop an efficient suboptimal resource allocation algorithm based on successive convex approximation to strike a balance between computational complexity and optimality.

\vspace*{-1mm}
\section{System Model}
In this section, we first present the considered MC UAV communication system model. Then, we discuss the solar energy harvesting model adopted for resource allocation design.

\vspace*{-2mm}
\subsection{Notation}%
We use boldface lower case letters to denote vectors.
$\mathbb{C}$ denotes the set of complex numbers; $\mathbb{R}^{N\times 1}$ denotes the set of all $N\times 1$ vectors with real entries; $\mathbb{R}^+$ denotes the set of non-negative real numbers; $\abs{\cdot}$ and $\norm{\cdot}$ denote the absolute value of a complex scalar and the Euclidean vector norm, respectively;
${\cal E}\{\cdot\}$ denotes statistical expectation;  the circularly symmetric complex Gaussian distribution with mean $\mu$ and variance $\sigma^2$ is denoted by ${\cal CN}(\mu,\sigma^2)$; and $\sim$ stands for ``distributed as"; $\nabla_{\mathbf{x}} f(\mathbf{x})$ denotes the gradient vector of function $f(\mathbf{x})$ whose components are the partial derivatives of $f(\mathbf{x})$.

\vspace*{-1mm}
\subsection{MC UAV Communication System Model}%
The considered MC UAV wireless communication system comprises one rotary-wing UAV-mounted transmitter \cite{Tech:LTEUAVQualcomm} and $K$ downlink (DL) users. The UAV-mounted transmitter and the DL users are single-antenna half-duplex devices, cf. Figure 1.
The UAV is equipped with solar panels which harvest solar energy and convert it to electrical energy. The harvested energy is used for providing communication services and powering the flight operation of the UAV.
The system bandwidth is divided into $N_{\mathrm{F}}$ orthogonal subcarriers. We assume that each subcarrier can be allocated to at most one user.
\begin{figure}
\centering\vspace*{-0mm}
\includegraphics[width=2.3in]{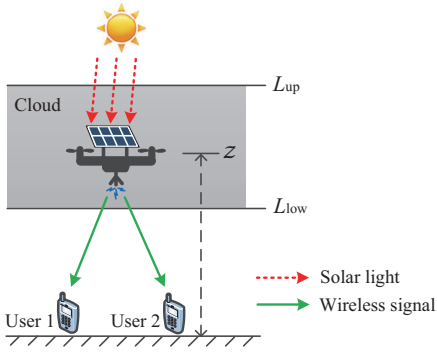}\vspace*{-3mm}
\caption{A solar powered MC UAV communication system with one UAV transmitter and $K=2$ downlink users.}
\label{fig:system_model}\vspace*{-5mm}
\end{figure}


In the considered system, the path loss of the communication link between the UAV and user $k \in \{1,\ldots,K\}$ is modeled as $\rho_k = \varrho\norm{\mathbf{r}-\mathbf{r}_k}^{-2}$,
where $\mathbf{r}=(x,y,z)$ and $\mathbf{r}_k=(x_k,y_k,0)$ represent the 3-D Cartesian coordinates of the UAV and user $k$, respectively. In particular, $(x,y)$ and $(x_k,y_k)$ are the horizontal coordinates of the UAV and user $k$, respectively, while $z$ denotes the altitude of the UAV. Besides, $\varrho=(\frac{c}{4\pi f_{\mathrm{0}}})^2$, where $c$ is the speed of light and $f_{\mathrm{0}}$ is the center frequency of the carrier signal.

Therefore, in each scheduling time slot, the received signal at downlink user $k$ on subcarrier $i\in\{1,\ldots, N_{\mathrm{F}}\}$ is given by \vspace*{-2mm}
\begin{eqnarray}
u_k^i = \frac{\sqrt{\varrho p_k^i}h_k^i }{\norm{\mathbf{r}-\mathbf{r}_k}} d_k^i + n_k^i,
\end{eqnarray}
where $d_k^i\in\mathbb{C}$ denotes the data symbol transmitted from the UAV to user $k$ on subcarrier $i$ and we assume ${\cal E}\{\abs{d_k^i}^2\}=1$ without loss of generality.  $p_k^i\in\mathbb{R}^+$ is the transmit power for the signal transmitted to user $k$ on subcarrier $i$. $h_k^i\in\mathbb{C}$ denotes the shadowing and small scale fading coefficient for the link between the UAV and user $k$ on subcarrier $i$ \cite{Hourani14Model}.
$n_k^i\sim{\cal CN}(0,\sigma_k^2)$ denotes the complex additive white Gaussian noise (AWGN) on subcarrier $i$ at user $k$. Besides, we assume that the channel state $h_k^i, \forall k,i,$ is perfectly known at the UAV to unveil a performance upper bound for MC UAV communication systems.

\vspace*{-2mm}
\subsection{Solar Energy Harvesting}
The considered MC UAV communication system is powered by the harvested solar energy. In general, the amount of harvested solar energy is affected by the atmospheric transmittance and clouds in the air \cite{duffie2013solar,kokhanovsky2004optical}. In particular, the atmospheric transmittance increases with the altitude, as higher altitudes lead to higher solar intensity such that more solar energy can be collected by the solar panels. The atmospheric transmittance for a given altitude is deterministic and has a value between zero and one which can be calculated by using the software tool LOWTRAN 7 \cite{kneizys1988users}. The atmospheric transmittance at altitude $z$ can be empirically approximated as follows \cite{Lee2017PathSolarUAV,larocca1975methods}: \vspace*{-1mm}
\begin{equation}
\phi(z)=\alpha - \beta e^{- z/\delta},
\end{equation}
where $\alpha$ is the maximum value of the atmospheric transmittance, $\beta$ is the extinction coefficient of the atmosphere, and $\delta$ is the scale height of the earth.
Besides, the solar intensity is reduced if there is a cloud between the sun and the solar panel. The attenuation of solar light passing through a cloud can be modeled as \cite{kokhanovsky2004optical}: \vspace*{-1mm}
\begin{equation}
\varphi(d^{\mathrm{cloud}})=e^{-\beta_c d^{\mathrm{cloud}}},
\end{equation}
where $\beta_c \ge 0$ denotes the absorption coefficient modeling the optical characteristics of the cloud and $d^{\mathrm{cloud}}$ denotes the distance that the solar light propagates through the cloud. Therefore, the electrical output power of the solar panels at altitude $z$ is modeled as \cite{duffie2013solar}\nocite{kokhanovsky2004optical}--\cite{Lee2017PathSolarUAV}: \vspace*{-2mm}
\begin{equation}\label{solar_power}
\hspace*{-2mm}P^{\mathrm{solar}}(z)\hspace*{-0.5mm}=
\hspace*{-0.5mm} \left\{
\begin{array}{lcl}
\hspace*{-2mm} \eta S G \phi(z)\varphi(0), & &\hspace*{-3mm} z \hspace*{-0.5mm} \ge \hspace*{-0.5mm} L_{\mathrm{up}},\\[-0mm]
\hspace*{-2mm} \eta S G \phi(z) \varphi(L_{\mathrm{up}}-z), & &\hspace*{-3mm} L_{\mathrm{low}} \hspace*{-0.5mm} \le \hspace*{-0.5mm} z \hspace*{-0.5mm} < \hspace*{-0.5mm} L_{\mathrm{up}},\\[-0mm]
\hspace*{-2mm} \eta S G \phi(z) \varphi(L_{\mathrm{up}}-L_{\mathrm{low}}), & &\hspace*{-3mm} z \hspace*{-0.5mm} < \hspace*{-0.5mm} L_{\mathrm{low}},
\end{array}\right.
\end{equation}
where $\eta$ and $S$ denote the energy harvesting efficiency and the area of the solar panels, respectively. Constant $G$ denotes the average solar radiation on earth. $L_{\mathrm{up}}$ and $L_{\mathrm{low}}$ are the altitudes of the upper and lower boundaries of the cloud, respectively.

\vspace*{-1mm}
\section{Problem Formulation and Solution}
In this section, after defining the adopted performance measure, we formulate the resource allocation problem. Then, we propose an iterative algorithm to solve the proposed problem.

\vspace*{-1mm}
\subsection{Achievable Data Rate}
Assuming subcarrier $i$ is allocated to user $k$, the achievable data rate on subcarrier $i$ is given by: \vspace*{-3mm}
\begin{equation}\label{rate_k}
U_k^i(p_k^i,s_k^i, \mathbf{r})=s_k^i \log_2 \Big( 1 + \frac{H_k^i p_k^i}{\norm{\mathbf{r}-\mathbf{r}_k}^2} \Big),
\end{equation}
where $H_k^i=\varrho\abs{h_k^i}^2/\sigma_k^2$. Variable $s_k^i \in \{0,1\}$ is the binary subcarrier allocation indicator. Specifically, $s_k^i = 1$ if user $k$ is allocated to subcarrier $i$, and $s_k^i = 0$, otherwise.

\vspace*{-2mm}
\subsection{Optimization Problem Formulation}
In this paper, we maximize the system sum throughput via optimizing the 3-D position and the power and subcarrier allocation of the UAV in a given time slot. The problem formulation is given as follows: \vspace*{-5mm}
\begin{eqnarray}
\label{prob}
&&\hspace*{-1mm} \underset{s_k^i, p_k^i,\mathbf{r}}{\maxo} \,\, \,\, \overset{N_{\mathrm{F}}}{\underset{i = 1}{\sum}} \overset{K}{\underset{k = 1}{\sum}} s_{k}^i \log_2 \Big( 1 + \frac{H_k^i p_k^i}{\norm{\mathbf{r}-\mathbf{r}_k}^2} \Big)  \\[-2mm]
\notag\mbox{s.t.}\hspace*{1mm}
&&\hspace*{-7mm}\mbox{C1: } \overset{N_{\mathrm{F}}}{\underset{i = 1}{\sum}} \overset{K}{\underset{k = 1}{\sum}}  s_{k}^i p_k^i \hspace*{-0.5mm}  + \hspace*{-0.5mm} P_{\mathrm{UAV}} \hspace*{-0.5mm} \le \hspace*{-0.5mm}  P^{\mathrm{solar}}(z), \hspace*{1.7mm}  \mbox{C2: } p_k^i \hspace*{-0.5mm}\ge \hspace*{-0.5mm} 0, \forall i,k, \notag\\[-1mm]
&&\hspace*{-7mm}\mbox{C3: } \overset{N_{\mathrm{F}}}{\underset{i = 1}{\sum}} \overset{K}{\underset{k = 1}{\sum}}  s_{k}^i p_k^i \le P_{\mathrm{max}},   \hspace*{10mm} \mbox{C4: } z_{\mathrm{min}} \le z \le z_{\mathrm{max}},   \notag \\[-3mm]
&&\hspace*{-7mm} \mbox{C5: } s_{k}^i \hspace*{-1mm}\in\hspace*{-1mm} \{0,1\}, \forall i,k,  \hspace*{17mm} \mbox{C6: } \overset{K}{\underset{k=1}{\sum}} s_{k}^i \hspace*{-1mm}\le\hspace*{-1mm} 1, \forall i. \notag
\end{eqnarray}
Constraint C1 is the power constraint of the UAV where constant $P_{\mathrm{UAV}}$ represents the power required for maintaining the operation of the UAV.  Constraint C2 is the non-negative transmit power constraint. $P_{\mathrm{max}}$ in constraint C3 denotes the maximum transmit power of the UAV-mounted transmitter as imposed by restrictions on the transmit spectrum mask to limit the amount of out-of-cell interference in the DL.
$z_{\mathrm{min}}$ and $z_{\mathrm{max}}$ in constraint C4 denote the minimum and the maximum flight altitude of the UAV imposed by regulation. Constraints C5 and C6 are imposed to guarantee that each subcarrier is allocated to at most one user.
For facilitating the presentation, we rewrite the power available from solar energy harvesting in \eqref{solar_power} as: \vspace*{-2mm}
\begin{equation}\label{solar_input}
\hspace*{-2mm}P^{\mathrm{solar}}(z)\hspace*{-0mm}= \hspace*{-0mm} \left \{
\begin{array}{lcl}
\hspace*{-2mm} A - B e^{-z/\delta}, & &\hspace*{-4mm} z \hspace*{-0.5mm} \ge \hspace*{-0.5mm}L_{\mathrm{up}},\\[-0mm]
\hspace*{-2mm} M(z)-B C_1 e^{(\beta_c-1/\delta) z}, & & \hspace*{-4mm} L_{\mathrm{low}} \hspace*{-0.5mm} \le \hspace*{-0.5mm} z \hspace*{-0.5mm} < \hspace*{-0.5mm} L_{\mathrm{up}},\\[-0mm]
\hspace*{-2mm} A C_2 - B C_2 e^{-z/\delta}, & &\hspace*{-4mm} z\hspace*{-0.5mm} < \hspace*{-0.5mm}L_{\mathrm{low}},
\end{array}\right.
\end{equation}
where $A=\eta S G \alpha$, $B=\eta S G \beta$, $C_1=e^{-\beta_c L_{\mathrm{up}}}$, $C_2=e^{-\beta_c (L_{\mathrm{up}}-L_{\mathrm{low}})}$, and $M(z)=A C_1 e^{\beta_c z}$. For the considered communication system, we note that there is a fundamental tradeoff between  harvesting solar energy and improving communication performance. In particular, the UAV can harvest more solar energy by climbing up to higher altitudes. However, flying at higher altitude leads to a larger path loss for the communication links between the UAV and the users which results in a degradation of the system performance.

The problem in \eqref{prob} is a mixed-integer non-convex problem due to the non-convex constraint C1, the binary constraint C5, and the non-convex objective function. In general, mixed-integer non-convex optimization problems cannot be solved optimally in a computationally efficient manner. Thus, in the next section, we propose a successive convex approximation based suboptimal scheme for the considered problem.

\vspace*{-1mm}
\subsection{Joint 3-D Position, Power, and Subcarrier Optimization}
In problem \eqref{prob}, the binary constraint C5 and constraint C6 are imposed to allocate at most one user to each subcarrier which is an obstacle for the design of a computationally efficient resource allocation algorithm. In this section, we transform problem \eqref{prob} into an equivalent form while relaxing constraints C5 and C6.
First, we temporarily assume that each subcarrier can be allocated to multiple users and introduce the following new utility function for user $k$ on subcarrier $i$: \vspace*{-2mm}
\begin{eqnarray}\label{equiv_rate}
\tilde{U}_k^i(\tilde{p}_k^i, \mathbf{r})\hspace*{-2mm} &=& \hspace*{-2mm} \log_2 \Bigg( 1 + \frac{\frac{H_k^i }{\norm{\mathbf{r}-\mathbf{r}_k}^2} \tilde{p}_k^i}{\frac{H_k^i }{\norm{\mathbf{r}-\mathbf{r}_k}^2} \sum_{m \ne k}^{K}  \tilde{p}_m^i + 1} \Bigg),
\end{eqnarray}
where $\tilde{p}_k^i \in \mathbb{R}^+$ denotes the transmit power for the signal transmitted to user $k$ on subcarrier $i$. In fact, \eqref{equiv_rate} represents the achievable rate of user $k$ on subcarrier $i$ where subcarrier $i$ is allocated to $K$ DL users and the term $\frac{H_k^i }{\norm{\mathbf{r}-\mathbf{r}_k}^2} \sum_{m \ne k}^{K} \tilde{p}_m^i$ in \eqref{equiv_rate} represents the multiuser interference at user $k$ from the $K-1$ co-channel users.
Then, adopting the utility function in \eqref{equiv_rate}, we formulate a modified optimization problem for maximizing the system sum throughput: \vspace*{-3mm}
\begin{eqnarray}
\label{equiv-prob}
&&\hspace*{-3mm} \underset{\tilde{\mathbf{p}}, \mathbf{r}}{\maxo} \,\, \,\, \overset{N_{\mathrm{F}}}{\underset{i = 1}{\sum}} \overset{K}{\underset{k = 1}{\sum}} \log_2 \Bigg( 1 + \frac{\frac{H_k^i }{\norm{\mathbf{r}-\mathbf{r}_k}^2} \tilde{p}_k^i}{ \frac{H_k^i }{\norm{\mathbf{r}-\mathbf{r}_k}^2} \sum_{m \ne k}^{K}  \tilde{p}_m^i + 1} \Bigg) \notag \\[-1mm]
\notag\mbox{s.t.}
&&\hspace*{-5mm}\widetilde{\mbox{C1}}\mbox{:  }\overset{N_{\mathrm{F}}}{\underset{i = 1}{\sum}} \overset{K}{\underset{k = 1}{\sum}}  \tilde{p}_k^i + P_{\mathrm{UAV}} \le  P^{\mathrm{solar}}(z),  \hspace*{2.5mm}  \\[-3mm]
&&\hspace*{-5mm} \widetilde{\mbox{C2}}\mbox{:  } \tilde{p}_k^i \ge 0, \,\, \forall i,k, \quad \widetilde{\mbox{C3}}\mbox{:  } \overset{N_{\mathrm{F}}}{\underset{i = 1}{\sum}} \overset{K}{\underset{k = 1}{\sum}}  \tilde{p}_k^i \le P_{\mathrm{max}}, \quad  \mbox{C4},
\end{eqnarray}
where $\tilde{\mathbf{p}}\in\mathbb{R}^{N_{\mathrm{F}}K \times1}$ is the collection of all $\tilde{p}_{k}^i$. We note that constraints $\widetilde{\mbox{C1}}$--$\widetilde{\mbox{C3}}$ and C4 in \eqref{equiv-prob} have the same physical meaning as the constraints C1--C4 in problem \eqref{prob}, respectively. Constraints C5 and C6 are not imposed in \eqref{equiv-prob} due to the modified subcarrier allocation strategy that allows the multiplexing of multiple users on each subcarrier.
We note that the problem formulations in \eqref{equiv-prob} and \eqref{prob} are equivalent when in \eqref{equiv-prob} on each subcarrier at most one of the powers $\tilde{p}_{k}^i$ is non-zero.
Now, we introduce the following theorem which reveals the equivalence between \eqref{equiv-prob} and \eqref{prob}.

\begin{table} \vspace*{-4mm}
\begin{algorithm} [H]                    
\caption{Successive Convex Approximation}    \vspace*{-0.2mm}      
\label{alg1}                           
\begin{algorithmic} [1]
\small          
\STATE Initialize the iteration index $j=1$ and initial point $\tilde{\mathbf{p}}^{(1)}$ and $\bm{\theta}^{(1)}$ \vspace*{-0.5mm}

\REPEAT \vspace*{-0.2mm}
\STATE For given $\tilde{\mathbf{p}}^{(j)}$ and $\bm{\theta}^{(j)}$, solve the convex problem in \eqref{prob-SCA} and store the resulting solution $\{\tilde{\mathbf{p}}$ and $\bm{\theta}\}$

\STATE Set $j=j+1$ and $\tilde{\mathbf{p}}^{(j)}=\tilde{\mathbf{p}}$ and $\bm{\theta}^{(j)}=\bm{\theta}$ \vspace*{-0.2mm}

\UNTIL convergence \vspace*{-0.2mm}

\STATE Obtain final resource allocation policy $\tilde{\mathbf{p}}^{*}=\tilde{\mathbf{p}}^{(j)}$, $\bm{\theta}^{*}=\bm{\theta}^{(j)}$
\end{algorithmic}
\end{algorithm}\vspace*{-9mm}
\end{table}

\begin{Thm}
The optimal subcarrier assignment strategy for maximizing the system sum throughput in \eqref{equiv-prob} assigns each subcarrier to the user with the best channel gain and no subcarrier is shared by multiple users.
\end{Thm}
\emph{\quad Proof: } The proof closely follows the proof in \cite[Appendix]{Jang03Transmit} and is omitted here due to the space limitation. \hfill\qed

Intuitively, if multiple users are activated on a subcarrier, the interference term $\frac{H_k^i }{\norm{\mathbf{r}-\mathbf{r}_k}^2} \sum_{m \ne k}^{K} \tilde{p}_m^i$ will severely degrade the system sum throughput. Therefore, problems \eqref{prob} and \eqref{equiv-prob} are equivalent in the sense that they yield the same optimal solution. Hence, we focus on the solution of problem \eqref{equiv-prob}. We note that the fractional term $\tilde{p}_{k}^i / \norm{\mathbf{r}-\mathbf{r}_k}^2$ in the objective function of \eqref{equiv-prob} is an obstacle to solving \eqref{equiv-prob} efficiently. We overcome this difficulty by rewriting \eqref{equiv-prob} in the following equivalent form: \vspace*{-2mm}
\begin{eqnarray}
\label{equiv-prob-2}
&&\hspace*{-5mm} \underset{\tilde{\mathbf{p}},\mathbf{r},\bm{\theta}}{\maxo} \,\, \,\, \overset{N_{\mathrm{F}}}{\underset{i = 1}{\sum}} \overset{K}{\underset{k = 1}{\sum}} \log_2 \Big( 1 + \frac{H_k^i \tilde{p}_k^i}{\sum_{m \ne k}^{K} H_k^i \tilde{p}_m^i + \theta_k} \Big)  \notag\\[-0mm]
&& \hspace*{-5mm}  \mbox{s.t.}
\hspace*{2mm}\widetilde{\mbox{C1}}\mbox{--}\widetilde{\mbox{C3}}, \mbox{C4}, \quad \mbox{C7: } \norm{\mathbf{r}-\mathbf{r}_k}^2 \le \theta_k,
\end{eqnarray}
where $\theta_k$ is an auxiliary variable and $\bm{\theta}\in\mathbb{R}^{K \times1}$ is the collection of all $\theta_{k}$. Now, the remaining non-convexity of problem \eqref{equiv-prob-2} is due to constraint $\widetilde{\mbox{C1}}$ and the objective function. We note that \eqref{equiv-prob-2} can be rewritten as a difference of convex programming problem \cite{dinh2010local}: \vspace*{-4mm}
\begin{eqnarray}
\label{prob-dc}
&&\hspace*{-5mm} \underset{\tilde{\mathbf{p}},\mathbf{r},\bm{\theta}}{\mino} \,\, \,\, -\sum_{i=1}^{{N_{\mathrm{F}}}}\sum_{k=1}^{K} \log_2 \Big(\sum_{m=1}^{K} H_k^i \tilde{p}_m^i + \theta_k \Big) - G(\tilde{\mathbf{p}},\bm{\theta})\notag \\[-1mm]
&&\hspace*{-7mm} \mbox{s.t.}
\hspace*{2mm}\widetilde{\mbox{C2}}, \widetilde{\mbox{C3}}, \mbox{C4}, \mbox{C7}, \quad \widetilde{\mbox{C1}}\mbox{:  }\overset{N_{\mathrm{F}}}{\underset{i = 1}{\sum}} \overset{K}{\underset{k = 1}{\sum}}  \tilde{p}_k^i \hspace*{-0.8mm}  +  \hspace*{-0.8mm} P_{\mathrm{UAV}} \hspace*{-0.8mm} \le \hspace*{-0.8mm} P^{\mathrm{solar}}(z),
\end{eqnarray}
where $G(\tilde{\mathbf{p}},\bm{\theta})\hspace*{-0mm}=\hspace*{-0mm}-\sum_{i=1}^{{N_{\mathrm{F}}}}\sum_{k=1}^{K} \log_2 \Big( \sum_{m \ne k}^{K} H_k^i \tilde{p}_m^i + \theta_k \Big).$
We note that the problems in \eqref{equiv-prob-2} and \eqref{prob-dc} are equivalent. We can obtain a locally optimal solution of \eqref{prob-dc} by applying successive convex approximation \cite{dinh2010local}. In particular, for any feasible point $\tilde{\mathbf{p}}^{(j)}$ and $\bm{\theta}^{(j)}$, we replace $G(\tilde{\mathbf{p}},\bm{\theta})$ and $P^{\mathrm{solar}}(z)$ with their global underestimations $\underline{G} (\tilde{\mathbf{p}},\bm{\theta}, \tilde{\mathbf{p}}^{(j)},\bm{\theta}^{(j)})$ and $\underline{P}^{\mathrm{solar}}(z)$, respectively, where \vspace*{-0mm}
\begin{eqnarray}\label{ineq1}
\hspace*{-2mm} \underline{G} (\tilde{\mathbf{p}},\bm{\theta}, \tilde{\mathbf{p}}^{(j)},\bm{\theta}^{(j)})
\hspace*{-2mm} &=& \hspace*{-2mm} G(\tilde{\mathbf{p}}^{(j)},\bm{\theta}^{(j)})  +  \nabla_{\tilde{\mathbf{p}}} G(\tilde{\mathbf{p}},\bm{\theta})(\tilde{\mathbf{p}} \hspace*{-0mm} - \hspace*{-0mm}\tilde{\mathbf{p}}^{(j)})  \notag \\[-1mm]
\hspace*{-2mm} &+& \hspace*{-2mm} \nabla_{\bm{\theta}} G(\tilde{\mathbf{p}},\bm{\theta})(\bm{\theta} \hspace*{-0mm} - \hspace*{-0mm}\bm{\theta}^{(j)}),
\end{eqnarray}
\begin{equation}
\hspace*{-2mm}\underline{P}^{\mathrm{solar}}(z)\hspace*{-1mm}=
\hspace*{-1mm} \left\{
\begin{array}{lcl}
\hspace*{-2mm} A - B e^{-z/\delta} , & &\hspace*{-6mm} z \hspace*{-1mm} \ge \hspace*{-1mm} L_{\mathrm{up}},\\[-0mm]
\hspace*{-2mm} \underline{M}(z,\hspace*{-0.5mm}z^{(j)}) \hspace*{-1mm} - \hspace*{-1mm} B C_1 e^{(\beta_c-1/\delta) z}, & &\hspace*{-6mm} L_{\mathrm{low}} \hspace*{-1mm} \le \hspace*{-1mm} z \hspace*{-1mm} < \hspace*{-1mm} L_{\mathrm{up}},\\[-0mm]
\hspace*{-2mm} A C_2 - B C_2 e^{-z/\delta}, & &\hspace*{-6mm} z \hspace*{-1mm} < \hspace*{-1mm} L_{\mathrm{low}}.
\end{array}\right.
\end{equation}\vspace*{-0mm}
Here, $\underline{M}(z,z^{(j)}) = A C_1 e^{\beta_c z^{(j)}} + A C_1 \beta_c e^{\beta_c z^{(j)}} (z-z^{(j)})$ is the global underestimation of $M(z)=A C_1 e^{\beta_c z}$ in \eqref{solar_input}.
Then, for any given $\tilde{\mathbf{p}}^{(j)}$ and $\bm{\theta}^{(j)}$, we can obtain a lower bound for \eqref{prob-dc} by solving the following optimization problem: \vspace*{-2mm}
\begin{eqnarray}
\label{prob-SCA}
&&\hspace*{-11mm} \underset{\tilde{\mathbf{p}},\mathbf{r},\bm{\theta}}{\mino}  -\hspace*{-0.6mm}\sum_{i=1}^{{N_{\mathrm{F}}}}\sum_{k=1}^{K} \log_2 \hspace*{-1mm} \Big( \hspace*{-1mm} \sum_{m=1}^{K} \hspace*{-1mm} H_k^i \tilde{p}_m^i \hspace*{-0.8mm} + \hspace*{-0.2mm} \theta_k \Big) \vspace*{-1.2mm} - \hspace*{-0.2mm} \underline{G} (\tilde{\mathbf{p}}, \hspace*{-0.2mm}\bm{\theta}, \hspace*{-0.mm}\tilde{\mathbf{p}}^{(j)}, \hspace*{-0.2mm}\bm{\theta}^{(j)}) \notag \\[-1mm]
\hspace*{-0mm} \mbox{s.t.}
&&\hspace*{-5mm}\widetilde{\mbox{C2}}, \widetilde{\mbox{C3}}, \mbox{C4}, \mbox{C7}, \hspace*{5mm} \widetilde{\mbox{C1}}\mbox{:  }\overset{N_{\mathrm{F}}}{\underset{i = 1}{\sum}} \overset{K}{\underset{k = 1}{\sum}}  \tilde{p}_k^i \hspace*{-0.8mm} + \hspace*{-0.8mm} P_{\mathrm{UAV}} \hspace*{-0.8mm}\le \hspace*{-0.8mm} \underline{P}^{\mathrm{solar}}(z),
\end{eqnarray}
Then, we successively tighten the obtained lower bound by applying the iterative algorithm summarized in \textbf{Algorithm 1}.
The proposed iterative algorithm converges to a locally optimal solution of \eqref{prob-dc} and has polynomial time computational complexity \cite{dinh2010local}.
We note that standard convex program solvers such as CVX \cite{website:CVX} can be used for efficiently solving the convex problem in \eqref{prob-SCA}.

Note that we can determine the subcarrier allocation policy from the obtained $\tilde{\mathbf{p}}^{*}$ in line 6 of \textbf{Algorithm 1}.
In particular, we note that $\tilde{p}_k^{i}$ is larger than zero only if DL user $k$ is allocated to subcarrier $i$. Thus, the subcarrier allocation policy is obtained as: $s_k^i = 1$ if $\tilde{p}_k^{i} > 0$, and $s_k^i = 0$, otherwise.


\vspace*{-1mm}
\section{Simulation Results}

In this section, we evaluate the system performance of the proposed scheme via simulations. The adopted simulation parameters are given in Table \ref{tab:parameters}.
We consider a single cell where the $K$ DL users are randomly and uniformly distributed within in the cell boundary of $1500$ meters and the entire service area is covered by clouds.
In each slot, the fading coefficients of the channels between the UAV and the DL users on each subcarrier are independent and identically distributed random variables following a Rician distribution with Rician factor $3$ dB. We obtained the results shown in this section by averaging over $5000$ realizations of multipath fading.
\begin{table}[t]\vspace*{-3mm}\caption{System parameters}\vspace*{-2mm}\label{tab:parameters} 
\newcommand{\tabincell}[2]{\begin{tabular}{@{}#1@{}}#2\end{tabular}}
\centering
\begin{tabular}{|l|l|}\hline
\hspace*{-1mm}Carrier center frequency and bandwidth & $2$ GHz and $5$ MHz \\
\hline
\hspace*{-1mm}Number and bandwidth of subcarriers & $64$  and $78$ kHz\\
\hline
\hspace*{-1mm}Parameters for atmospheric transmittance, $\alpha$, $\beta$   & $0.8978$, $0.2804$ \cite{duffie2013solar} \\
\hline
\hspace*{-1mm}Average solar radiation and scale height, $G$, $\delta$  & $1367$ $\mathrm{W}\hspace*{-0.4mm}/\hspace*{-0.4mm}\mathrm{m}^2$, $8000$ m \\
\hline
\hspace*{-1mm}Altitude of cloud, $L_{\mathrm{low}}$ and $L_{\mathrm{up}}$  & $700$ m and $1400$ m \cite{kokhanovsky2004optical} \\
\hline
\hspace*{-1mm}Absorption coefficient of cloud, $\beta_c$ &  $0.01$ \cite{kokhanovsky2004optical} \\
\hline
\hspace*{-1mm}Altitude limitation for UAV, $z_{\mathrm{min}}$ and $z_{\mathrm{max}}$ & $100$ $\mathrm{m}$ and $1500$ $\mathrm{m}$  \\
\hline
\hspace*{-1mm}Efficiency of solar panels, $\eta$ & $0.4$   \\
\hline

\hspace*{-1mm}Receiver noise power, $\sigma_{k}^2$   &   \mbox{$-110$ dBm}  \\
\hline
\hspace*{-1mm}Power requirement for operating UAV, $P_{\mathrm{UAV}}$ &  \mbox{$200$ W}   \\
\hline
\end{tabular}
\vspace*{-0mm}
\end{table}
Besides, we also consider the performance of two baseline schemes for comparison.
For baseline scheme 1, we set $(x,y)\hspace*{-1mm}=\hspace*{-1mm}(0,0)$, i.e., the origin of the cell, and then jointly optimize the flight altitude $z$, power $p_k^i$, and subcarrier allocation $s_k^i$ of the UAV communication system.
For baseline scheme 2, the user on each subcarrier is selected randomly and we optimize the 3-D position of the UAV and the power allocated to the users.

\begin{figure}[t]
 \centering\vspace*{-0mm}
\includegraphics[width=3.2in]{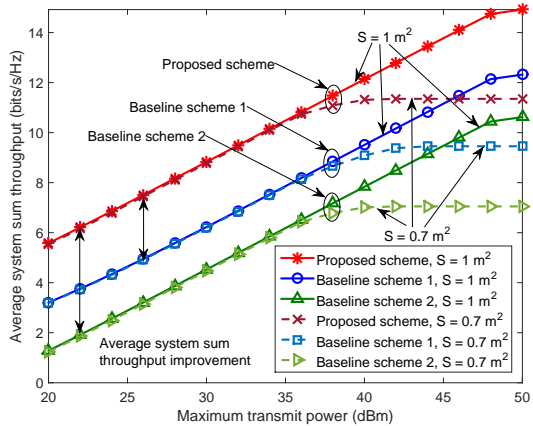} \vspace*{-3mm}
\caption{Average system sum throughput (bits/s/Hz) versus the maximum transmit power of the UAV (dBm), $P_{\mathrm{max}}$, for different resource allocation schemes and $K=3$. } \label{fig:wsr_vs_power}\vspace*{-4mm}
\end{figure}

In Figure \ref{fig:wsr_vs_power}, we investigate the average system sum throughput versus the maximum transmit power at the UAV, $P_{\mathrm{max}}$, for $K=3$ DL users and different solar panel sizes $S$. The average system throughputs of the proposed scheme and all baseline schemes increase monotonically with the maximum transmit power $P_{\mathrm{max}}$.
In fact, for the proposed scheme and the baseline schemes, the UAV can fly to higher altitudes to harvest more solar energy when the maximum transmit power $P_{\mathrm{max}}$ increases. Thus, the proposed scheme and the baseline schemes can effectively exploit the increased transmit power budget to improve the received signal-to-interference-plus-noise ratio (SINR) at the users.
Besides, for a smaller value of $S$, there is a diminishing return in the average system sum throughput for the proposed scheme and all baseline schemes when $P_{\mathrm{max}}$ exceeds $35$ dBm and the average system sum throughput approaches a constant in the large $P_{\mathrm{max}}$ regime.
In fact, since the output power of the solar panels is smaller for smaller solar panel sizes, a UAV equipped with a smaller solar panel has to fly at a higher altitude to harvest the same amount of solar energy as a UAV equipped with a larger solar panel, which leads to a severe path loss for air-to-ground communications and causes a performance degradation.
Furthermore, as can be observed, the proposed scheme achieves a considerably higher average system sum throughput than baseline schemes $1$ and $2$ due to the joint optimization of the 3-D position and the power and subcarrier allocation.
In particular, for baseline scheme $1$, the horizontal coordinates of the UAV $(x,y)$ are fixed leading to fewer available degrees of freedom for improving the average system sum throughput. For baseline scheme $2$, although the adopted random subcarrier allocation policy provides fairness, it causes a poor utilization of the system resources.

\begin{figure}[t]
 \centering\vspace*{-2mm}
\includegraphics[width=3.2in]{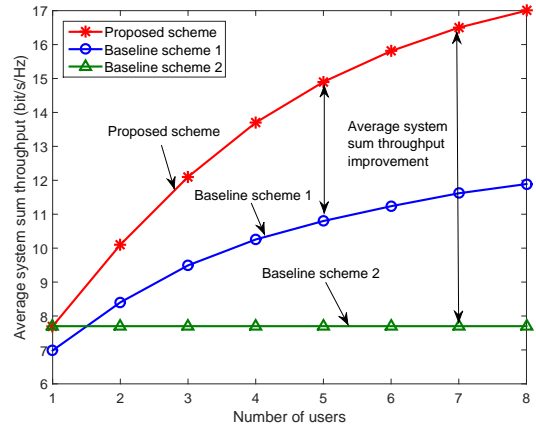} \vspace*{-3mm}
\caption{Average system sum throughput (bits/s/Hz) versus  the number of users for different resource allocation schemes with $P_{\mathrm{max}}\hspace*{-0.5mm}=\hspace*{-0.5mm}40$ dBm and $S=1$ ${\mathrm{m}^2}$. }\label{fig:wsr_vs_usernum}\vspace*{-4mm}
\end{figure}

In Figure \ref{fig:wsr_vs_usernum}, we study the average system sum throughput versus the number of DL users for $P_{\mathrm{max}}\hspace*{-0.5mm}=\hspace*{-0.5mm}40$ dBm and $S \hspace*{-0.5mm} =\hspace*{-0.5mm} 1$ $\mathrm{m}^2$.
As can be observed, the average system sum throughput for the proposed scheme and baseline scheme $1$ increase with the number of users since these schemes are able to exploit multiuser diversity.
The performance of baseline scheme $2$ is independent of the number of users since it employs a random subcarrier allocation policy.
Moreover, it can be observed from Figure \ref{fig:wsr_vs_usernum} that the average system sum throughput of the proposed scheme grows faster with the number of users than that of baseline scheme $1$.
In fact, for baseline scheme $1$, the UAV cannot adjust its horizontal coordinates $(x,y)$ according to the locations of the users
which limits its capability to exploit the multiuser diversity introduced by the different locations of the users.

\vspace*{-2mm}
\section{Conclusion}
In this paper, we studied the joint optimization of the 3-D position and the power and subcarrier allocation for solar powered MC UAV communication systems. The objective of the resulting mixed-integer non-convex optimization problem was the maximization of the system sum throughput. A suboptimal resource allocation algorithm design based on successive convex approximation was proposed. Simulation results unveiled that the proposed scheme for solar powered UAV systems achieves a significant improvement in system performance compared to two baseline schemes.

\vfill\pagebreak
\bibliographystyle{IEEEtran}
\bibliography{UAV_solar_MC}

\end{document}